\documentclass[prd,aps,preprint,amsfonts,showpacs]{revtex4}
\usepackage{axodraw}
\newcommand{\ar}{\arrowvert}
\newcommand{\ra}{\rangle}
\newcommand{\la}{\langle}

\newcommand{\ov}{\overline}
\newcommand{\cd}{\! \cdot \!}
\newcommand{\be}{\begin{equation}}
\newcommand{\ee}{\end{equation}}
\newcommand{\ba}{\begin{eqnarray}}
\newcommand{\ea}{\end{eqnarray}}
\input{psfig.sty}
\usepackage{graphicx}
\usepackage{dcolumn}
\usepackage{graphics}
\begin{document}

\title{Is the $\Theta ^+$ a $K \pi N$ bound state?\\}

\author{Felipe J. Llanes-Estrada } \email{fllanes@fis.ucm.es}
\affiliation{Departamento de F\'{\i}sica Te\'orica I,  Universidad
Complutense, 28040 Madrid, Spain}
\author{E. Oset} \email{oset@ific.uv.es} 
\author{V. Mateu}
\affiliation{Departamento de F\'{\i}sica Te\'orica and IFIC,
Centro Mixto Universidad de Valencia-CSIC,
Institutos de Investigaci\'on de Paterna, Aptd. 22085, 46071
Valencia, Spain.}

\date{\today}

\date{\today}

\begin{abstract}
 Following a recent suggestion that the $\Theta ^+$ could be a
 $K \pi N$ bound state we perform an investigation under the 
 light of the meson meson and meson baryon dynamics provided by the chiral
 Lagrangians and using methods currently employed to dynamically generate 
meson and baryon resonances by means of unitary extensions of 
chiral perturbation theory.  We consider two body and three body forces 
and examine the possibility of a bound state below the three 
particle pion-kaon-nucleon and above the
kaon-nucleon thresholds.  
Although we find indeed an attractive interaction in the case of isospin 
I=0 and spin-parity $1/2^+$, the interaction is too weak to bind the 
system. If we arbitrarily add to the physically motivated potential
the needed strength to bind the system and with such 
strong attraction evaluate the decay width into $K N$, this turns out 
to be small. A discussion on further  work in this direction is done.
\end{abstract}

\maketitle

\section{Introduction}

A recent experiment at SPring-8/Osaka~\cite{Nakano:2003qx}
has found a clear signal for an $S=+1$, positive charge resonance
around 1540 MeV, confirmed
 by the DIANA collaboration at
ITEP~\cite{Barmin:2003vv}, CLAS at Jefferson
Lab.~\cite{Stepanyan:2003qr} and SAPHIR at ELSA~\cite{Barth:2003es}. 
The resonance has explicit exotic flavor quantum numbers given
the decay final states $K^0 p$ and $K^+ n$.
Its width is also intriguingly narrow, less than 20 $MeV$ by present
experimental bounds. 
A  state with these characteristics was originally predicted
by Diakonov {\em et~al.} in Ref.~\cite{Diakonov:1997mm}, and since the
experimental observation a large number of theoretical papers have 
appeared with different suggestions as to the nature of the state and 
possible partners
\cite{Capstick:2003iq,Stancu:2003if,
Karliner:2003sy,hosaka,Jaffe:2003sg,Zhu:2003ba,
Glozman:2003sy,Kim:2003ay,Matheus:2003xr,
Cohen:2003yi,Cohen:2003nk,Csikor:2003ng,Liu:2003rh,Nam:2003uf,Liu:2003zi,
Huang:2003}.
 Most of the works look at the quark
structure of what is being called the pentaquark, since a standard three 
quark Fock space assignment is not allowed.
The parity of this candidate state is
as of yet undetermined \cite{hyodo}, 
and whereas quark model calculations in the ground 
state \cite{Strottman:qu,carlson,williams} 
assign to it negative parity, 
positive parity is predicted in the Skyrme model \cite{Diakonov:1997mm} 
requiring a p-wave in the quark model \cite{Stancu:2003if},\cite{carlson2}.

Yet, at a time when many  low energy baryonic resonances are being
dynamically generated as meson baryon quasibound states within chiral 
unitary
approaches \cite{Kaiser:1995eg,Oset:1997it,Oller:2000fj,Inoue:2001ip,
Garcia-Recio:2002td,Jido:2003cb,Garcia-Recio:2003ks,Nieves:2001wt} it 
looks tempting to
investigate the
possibility of this state being a quasibound state of a meson and a baryon 
or two  mesons and a baryon.  Its nature as a $K N$ s-wave state is easily
ruled 
out since the interaction is repulsive. This is in general the case for 
scattering with exotic quantum numbers (not attainable with three
quarks) which also explains the repulsive core nucleon-nucleon 
interaction \cite{ribeiro}. Indeed the known kaon-nucleon phase 
shifts seem difficult to reconcile with the existence of a broad
$\Theta^+$ resonance, although a narrow one is not excluded 
\cite{KNshifts}. $KN$ in a p-wave, which is attractive, is too weak to 
bind. The next logical possibility is to 
consider a quasibound state of $K 
\pi N$, which in s-wave would naturally correspond to spin-parity 
$1/2^+$, the quantum numbers suggested in
\cite{Diakonov:1997mm}.
Such an idea has already been put forward in \cite{Bicudo:2003rw} where a  
study of the interaction of the three body system is conducted in the 
context of chiral quark models, which 
suggests  that it
is not easy to bind the system although one cannot rule it out completely.
Similar ideas have been exploited in the past \cite{Longacre:1990uc}
to describe the $f_1 \ (1420)$
meson, then named E(1420), as a $K K \pi$ molecule bound by color singlet
exchanges.

 In the present work we  further investigate in this direction and for
 this we use the meson meson and meson baryon interactions generated by 
the chiral Lagrangians and apply techniques of unitarized chiral perturbation
 theory which have been used in the dynamical generation of the low lying
 baryonic resonances.

\section{A $\kappa N$ state?} 

Upon considering the possible structure of $\Theta^+$ 
we are guided by the  experimental observation
\cite{Stepanyan:2003qr} that the
state is not produced in the $K^+ p$ final state. This would
rule out the possibility of the $\Theta$ state having 
isospin I=1. Then we accept the $\Theta^+$ to be an I=0 state. 
As we couple a pion and a kaon to the nucleon to form such state,
a consequence is that the $K \pi$ substate must combine to I=1/2 and 
not I=3/2.
 This is also welcome dynamically since the s-wave $K \pi$ interaction in
I=1/2 is attractive (in I=3/2 repulsive) \cite{Oller:1998hw0}.
The attractive interaction in I=1/2 is very strong and gives
rise to the dynamical generation of the scalar $\kappa$ resonance around 
850 MeV and with a large width \cite{Oller:1998hw0}.

One might be tempted to consider the $\Theta^+$ state as a quasibound 
$\kappa N$ state. However the $\Theta^+$ state would then be bound by
about 200 MeV, apparently too large an amount. But recall that 
the large width of the $\kappa$ (around 400 MeV) allows $\kappa$ 
strength at lower
energies and the large binding becomes  more relative. One might next
question that, with such a large width of the $\kappa$, the $\Theta^+$
could not be so narrow as experimentally reported.
However, this large $\kappa$ width is no problem since in our scenario 
it would arise from $K\pi$ decay, but now the $K \pi N$ decay of the 
$\Theta^+$ is forbidden as the $\Theta^+$ mass is below the $K \pi N$  
threshold.

One might hesitate to call the possible theoretical $\Theta^+$ state a
$\kappa N$ quasibound state because of the large gap to the
nominal $\kappa N$ mass. The name though is not relevant here and we can 
opt by calling it simply a $K \pi N$ state, but the fact is that the
$K\pi$ system is strongly correlated even at these lower energies, and
since 
this favours the binding of the $K \pi N$ state we shall take it into
account.

\subsection{$K \pi$ Scattering Matrix.}
We begin by refreshing how the $\kappa$ can be generated in the Bethe 
Salpeter approach used in \cite{Oller:1997ti} to generate the $\sigma$, 
$f_0(980)$ and $a_0(980)$ scalar resonances. From the lowest order
ChPT Lagrangian \cite{Gasser:1985gg}
one takes the $K \pi$ amplitude which serves as kernel, V, of
the Bethe Salpeter equation  (here the Lippman-Schwinger equation
with relativistic meson propagator)

\be \label{kappaLippman}
t_{K \pi} = V_{K \pi}+V_{K \pi} G_{mm} t_{K \pi}
\ee

where $V_{K \pi}$ for I=1/2 in s-wave, which we call from now on $t_{mm}$, 
is given by

\be \label{pionkaonvertex}
t_{mm} =
\begin{picture}(50,50)(-20,0)
\DashArrowLine(0,0)(25,25){7}   \put(0,10){$K$}
\DashArrowLine(0,50)(25,25){3}   \put(0,32){$p$}
\DashArrowLine(25,25)(50,0){7} \put(45,10){$K'$}
\DashArrowLine(25,25)(50,50){3} \put(45,32){$p'$}
\end{picture}
\ee
yielding (in an s-wave)
\be
\la I=1/2 \ I_3 \ar t_{mm} \ar I=1/2 \ I_3 \ra =
\frac{4m_\pi^2+4m_K^2-4 s + \frac{3\lambda}{2s}}{4f^2}
\ee
where $f\simeq 100 MeV$ is the meson decay constant, which we take as an 
average between $f_\pi$ and $f_K$, $s$ the
Mandelstam variable, and $\lambda(m_\pi,m_K,\sqrt{s})=(m_\pi^4+m_K^4+
s^2-2 (m_\pi^2 m_K^2 + m_\pi^2 s + m_K^2 s) )$  K\"allen's
function.

Also in eq. (\ref{kappaLippman}) $G_{mm}$ is the two meson loop 
function defined in \cite{Oller:1997ti}  
and regularized with a three-momentum cutoff of 850 $MeV$ 
(which produces satisfactory fits to the $\pi K$ scattering phase shift in 
the $\kappa$  and also in the $\pi \pi$ $\sigma$ channels, not 
shown), 
\be \label{mesonloop}
G_{mm}=\int_0^\Lambda  \frac{q^2dq}{4\pi^2}
\frac{\omega_K+\omega_\pi}{\omega_K \omega_\pi} \frac{1}{(\sqrt{s} +
\omega_\pi + \omega_K)(\sqrt{s}-\omega_\pi -\omega_K + i\epsilon)} 
\ee
\be \label{lasomegas}
\omega_K=\sqrt{m_K^2+q^2}, \ \omega_\pi=\sqrt{m_\pi^2+q^2}
\ee
and $V$, $t$
factorize in eq. (\ref{kappaLippman}) with their on shell value as
discussed in \cite{Oller:1997ti}.
This simply means that one takes $p_i^2$=$m_i^2$ in the expressions of
the $K \pi$ kernel. Note that $t_{mm}$ is attractive in the $\kappa$ 
channel.

Eq. (\ref{kappaLippman}), which we numerically solve,
resums the $\pi K$  scattering perturbation series 

\be \label{kappaprop}
\begin{picture}(50,50)(0,0)
\Boxc(25,25)(10,20)
\DashArrowLine(0,0)(20,20){7}  \DashArrowLine(30,20)(50,0){7}  
\DashArrowLine(0,50)(20,30){3} \DashArrowLine(30,30)(50,50){3}
\end{picture}
=
\begin{picture}(50,50)(0,0)
\DashArrowLine(0,0)(25,25){7}   \put(0,12){$K$}
\DashArrowLine(0,50)(25,25){3}   \put(0,32){$p$}
\DashArrowLine(25,25)(50,0){7} \put(47,12){$K'$}
\DashArrowLine(25,25)(50,50){3} \put(47,32){$p'$}
\end{picture}
\ \ + .\ .\ . \ \ + \ \ 
\begin{picture}(200,40)(0,0)
\DashArrowLine(0,0)(20,20){7}   \put(-5,5){$K$}
\DashArrowLine(0,40)(20,20){3}   \put(0,25){$p$}
\DashArrowLine(180,20)(200,0){7} \put(195,5){$K'$}
\DashArrowLine(180,20)(200,40){3} \put(195,25){$p'$}
\DashArrowArc(40,20)(20,180,0){7} \DashArrowArc(80,20)(20,180,0){7}
\DashArrowArc(120,20)(20,180,0){7} \DashArrowArc(160,20)(20,180,0){7}
\DashArrowArc(40,20)(20,0,180){3} \DashArrowArc(80,20)(20,0,190){3}
\DashArrowArc(120,20)(20,0,180){3} \DashArrowArc(160,20)(20,0,180){3}
\end{picture}
+ . \ . \ .
\ee

The $\kappa$ state appears then as a pole of the $t_{K \pi}$ matrix in 
the complex plane.

\subsection{$N \kappa$ Scattering.}

In order to determine the possible $\Theta^+$ state we search for poles of
the $K\pi N \to K\pi N$ scattering matrix. To such point we construct the
series of diagrams

\be \label{Thetaprop}
\begin{picture}(50,50)(0,0)
\SetWidth{2}
\ArrowLine(0,0)(25,0) \ArrowLine(25,0)(50,0)
\SetWidth{0.5}
\DashArrowLine(0,15)(25,0){3} \DashArrowLine(25,0)(50,15){3}
\DashArrowLine(0,30)(25,0){7} \DashArrowLine(25,0)(50,30){7}
\end{picture}
\ \ +\ \ 
\begin{picture}(80,50)(0,0)
\SetWidth{2}
\ArrowLine(0,0)(25,0) \ArrowLine(25,0)(49,0) \ArrowLine(49,0)(74,0)
\SetWidth{0.5}
\DashArrowLine(0,15)(25,0){3} 
\DashArrowLine(49,0)(74,15){3}
\DashArrowLine(0,30)(25,0){7} \DashArrowLine(49,0)(74,30){7}
\DashArrowArcn(37,0)(12,180,0){3}
\DashArrowArcn(37,-10)(18,140,40){2}
\end{picture}
\ \ + \ \ 
\begin{picture}(80,50)(0,0)   
\SetWidth{2}
\ArrowLine(0,0)(25,0) \ArrowLine(25,0)(49,0) \ArrowLine(49,0)(74,0)
\SetWidth{0.5}
\DashArrowLine(0,15)(25,0){3}
\DashArrowLine(49,0)(74,15){3}
\DashArrowLine(0,30)(25,0){7} \DashArrowLine(49,0)(74,30){7}
\Oval(30,9)(10,4)(135) \Oval(45,9)(10,4)(45)
\end{picture}
+ .\ .\ . +
\begin{picture}(130,50)(0,0)
\SetWidth{2}
\ArrowLine(0,0)(25,0) \ArrowLine(25,0)(49,0) \ArrowLine(49,0)(99,0) 
\ArrowLine(99,0)(125,0)
\SetWidth{0.5}
\DashArrowLine(0,15)(25,0){3} \DashArrowLine(99,0)(125,15){3}
\DashArrowLine(0,30)(25,0){7} \DashArrowLine(99,0)(125,30){7}
\Oval(30,9)(10,4)(135) \Oval(45,9)(10,4)(45)
\Oval(60,9)(10,4)(135) \Oval(94,9)(10,4)(45) \Oval(77,15)(10,4)(90) 
\end{picture}
+ . \ . \ .
\ee
where we account explicitly for the $K\pi$ interaction by constructing
correlated $K\pi$ pairs and letting the intermediate $K\pi$ and nucleon 
propagate. 
This requires a kernel for the two meson-nucleon interaction which we
now address. The $K\pi$ correlation in the external legs is dispensable   
for the purpose of finding poles of the $t$ matrix.

We formulate the meson-baryon lagrangian in terms of
the SU(3) matrices, $B$, $\Gamma_\mu$, $u$ and the implicit meson matrix 
$\Phi$ standard in ChPT \cite{Meissner:1993ah,
Pich:1995bw,Ecker:1995gg,Bernard:1995dp},
\ba \label{lagrangian} \nonumber
{\mathcal L}= {\rm Tr} \left( \ov{B} i\gamma^\mu \nabla_\mu B \right)
- M_B {\rm Tr}\left(\ov{B}B  \right) + \\
+ \frac{1}{2} D {\rm Tr} \left( \ov{B} \gamma^\mu \gamma_5 \left\{
u_\mu,B\right\}   \right) +
\frac{1}{2} F {\rm Tr} \left( \ov{B} \gamma^\mu \gamma_5 \left[
u_\mu, B \right] \right)
\ea
\ba \nonumber
\nabla_\mu B= \partial_\mu B + [\Gamma_\mu,B]
\\
\Gamma_\mu=\frac{1}{2}(u^\dagger\partial_\mu u+u\partial_\mu u^\dagger) 
\ea
with the definitions in \cite{Meissner:1993ah,
Pich:1995bw,Ecker:1995gg,Bernard:1995dp}.

First there is a contact three body force simultaneously involving 
the pion, kaon and nucleon, which can be derived from the meson-
baryon Lagrangian (\ref{lagrangian}) term containing  $\Gamma_{\mu}$.

\begin{equation} \label{kappaNvertex1}
t_{mB}^s = \ \
\begin{picture}(100,100)(0,0)
\SetWidth{2}
\ArrowLine(0,0)(50,0)   \put(0,5){$N, M_p$}
\ArrowLine(50,0)(100,0) \put(95,5){$M_p$}
\SetWidth{0.5}
\DashArrowLine(15,50)(50,0){7}   \put(0,55){$K,(K)$}
\DashArrowLine(5,30)(50,0){3}   \put(-5,30){$\pi,(p)$}
\DashArrowLine(50,0)(100,50){7} \put(95,55){$(K')$}
\DashArrowLine(50,0)(100,25){3} \put(95,30){$(p')$}
\Vertex(50,0){2}
\end{picture}
\end{equation}

We now show that a nucleon, kaon and pion see an attractive interaction
in an isospin zero state through this contact potential. 
By taking the isospin I=1/2 $\kappa$ states
\ba
\nonumber \kappa^0 = \frac{1}{\sqrt{3}} \ \ar \pi^0 \ K^0 \ra -
\sqrt{\frac{2}{3}} \ \ar \pi^- \ K^+ \ra \\
\kappa^+ = -\sqrt{\frac{2}{3}} \ \ar \pi^+ \ K^0 \ra -
\frac{1}{\sqrt{3}} \ \ar \pi^0 \ K^+\ra \ .
\ea
and combining them with the nucleon, also isospin $1/2$, we generate 
I=0,1 states
\ba \nonumber
\Theta^0 = \ar I=0 \ I_3=0 \ra = \frac{1}{\sqrt{2}} \ \left(
\ar P \ \kappa^0 \ra - \ar n \ \kappa^+ \ra \right) \\
\Theta^1 = \ar I=1 \ I_3=0 \ra = \frac{1}{\sqrt{2}} \ \left(
\ar P \ \kappa^0 \ra + \ar n \ \kappa^+ \ra \right) \ .
\ea
which diagonalize the scattering matrix associated to $t_{mB}$ 
\ba \label{diagtmatrix} \nonumber
\la \Theta^1 \ar t_{mB}^s \ar \Theta^1 \ra = -\frac{1}{144 f^4}
\left( -4 (\not K+ \not K') - 11 (\not p + \not p')\right)  \\
\la \Theta^0 \ar t_{mB}^s \ar \Theta^0 \ra = -\frac{21}{144 f^4}
\left(  (\not K + \not K')-  (\not p +\not p') \right)
\ea
where for a near-threshold study we will perform the usual 
non-relativistic approximation
$\bar{u}\gamma^\mu k_\mu u$ = $k^0$. Since the $K \pi N$ system is bound
by about 30 MeV one can take for a first test $k^0$, $p^0$ as the masses 
of
the $K$ and $\pi$
respectively and one sees that the interaction in the I=0 channel is
attractive, while in the I=1 channel is repulsive. 
This would give chances to the $\kappa N$ $t$-matrix to develop a pole in 
the bound region, but rules out the I=1 state. 

The series  (\ref{Thetaprop}) might lead to a bound state of
$\kappa N$ which would not decay since the only intermediate channel
is made out of $K \pi N$ with mass above the available energy. 

The decay into $K N$ observed experimentally can
be taken into account by explicitly allowing for an intermediate
state provided by the p-wave interaction vertices from (\ref{lagrangian}),
through the diagram
\be \label{kappaNvertex2}
t_{mB}^p = \ \
\begin{picture}(200,80)(0,0)
\SetWidth{2}
\ArrowLine(0,0)(68,0) \ArrowLine(68,0)(134,0) \ArrowLine(134,0)(200,0)
\SetWidth{0.5}
\DashArrowLine(0,80)(48,65){7}  \DashArrowLine(48,65)(154,65){7}
\DashArrowLine(154,65)(200,80){7} 
\DashArrowLine(0,50)(48,65){3} \DashArrowLine(48,65)(68,0){3}
\DashArrowLine(134,0)(154,65){3} \DashArrowLine(154,65)(200,50){3}
\put(65,30){$\pi$} \put(128,30){$\pi$}\put(97,69){$K$}
\put(-5,62){$(P^0)$} \put(190,62){$(P^0)$}
\end{picture} \ .
\ee
The evaluation of this diagram requires the extra $\pi NN$ Yukawa 
vertex, which one generates from the $D$, $F$ terms of the Lagrangian 
(\ref{lagrangian}) and to which we attach the commonly used $\pi NN$ 
monopole form  factor to account for the nucleon's finite size
with a scale $\Lambda=1 \ GeV$
\be
t_{\pi N}^{I_j} = i \left( \frac{G_A}{2f} \right) \vec{\sigma}\cd \vec{q}
F(\ar \vec{q} \ar^2) \la N \ar \tau^{I_j} \ar N' \ra
\ee
with $G_A=D+F=1.26$,
$$
F(\ar \vec{q} \ar^2) = \frac{\Lambda^2}{\Lambda^2+\ar\vec{q}\ar^2} 
$$
The isospin factor for (\ref{kappaNvertex2}) turns out to be 3 for $I=0$ 
and 1/3 for $I=1$. As we shall see, this diagram provides some 
attraction at low energies, but in the $I=1$ case the relative factor of 
1/9 makes it
negligible compared with the repulsion generated by eq. 
(\ref{kappaNvertex1}).
The evaluation of the customary pole integrals over 
$q^0$ in (\ref{kappaNvertex2}) leads to 
\ba \label{supertmb2} \nonumber
t_{mB}^p = 3 t_{mm}^2 \int \frac{d^3q}{(2\pi)^3} 
\left(\frac{G_A}{2f}\right)^2 
\ar \vec{q}\ar^2 F(q)^2 \frac{M_N}{E_N}
\frac{1}{p^0+P^0-\omega_K-E_N+i\epsilon} \\ \nonumber
\frac{1}{4\omega_k \omega_\pi^3(E_N-p^0+\omega_\pi)^2 
(P_0^2-(\omega_\pi +\omega_K)^2)^2} 
\\ \nonumber \left\{
(E_N-p_0)^2(-P_0^2\omega_K + 
(\omega_K+\omega_\pi)^2(\omega_K+2\omega_\pi))
\right. \\ \nonumber 
+(E_N-p^0)(P^{0 \ 3}\omega_K -P_0^2\omega_K(\omega_K+2\omega_\pi)+
(\omega_K^2+3\omega_\pi\omega_K+2\omega_\pi^2)^2 -P^0\omega_K
(\omega_K^2+4\omega_\pi \omega_K +3\omega_\pi^2) )
\\ \nonumber \left.
+2\omega_\pi(P^{0\ 3} \omega_K-P_0^2\omega_K(\omega_K+\omega_\pi) +
(\omega_K+\omega_\pi)^4 -P^0\omega_K(\omega_K^2+3\omega_\pi\omega_K 
+2\omega_\pi^2)) 
\right\}
\\
\ea
with $\omega_K$ and $\omega_\pi$ as in eq. (\ref{lasomegas}),
$E_N=\sqrt{M_N^2+q^2}$,  $p^0=M_N$ 
the nucleon mass (incoming and outgoing energies) and $P^0$ the incoming 
pion-kaon system energy (masses minus possible binding energy).

Through the remaining $N K$ propagator in the integral,
eq. (\ref{supertmb2}) 
generates a real part from the principal value and an imaginary part
corresponding to placing the intermediate $K$ and $N$ on shell. This would
account for the decay of the $\Theta^+$ state into $K N$.

\subsection{Sequential two body contributions}

The existence of diagram (\ref{kappaNvertex2}) above can be interpreted 
as having $\pi K$ interaction followed by $\pi N$ interaction in p-wave. 
One of course can also consider this latter interaction in s-wave using
the same  Lagrangian (\ref{lagrangian}) with two meson fields, as in
\be \label{kappaNvertex3}
t_{mB}^{s'} = \ \
\begin{picture}(200,80)(0,0)
\SetWidth{2}
\ArrowLine(0,0)(100,0) \ArrowLine(100,0)(200,0)
\SetWidth{0.5}
\DashArrowLine(0,80)(48,65){7}  \DashArrowLine(48,65)(154,65){7}
\DashArrowLine(154,65)(200,80){7}
\DashArrowLine(0,50)(48,65){3} \DashArrowLine(48,65)(100,0){3}
\DashArrowLine(100,0)(154,65){3} \DashArrowLine(154,65)(200,50){3}
\put(-5,62){$(P^0)$} \put(190,62){$(P^0)$}
\end{picture} \ .
\ee
There is also a novelty with respect to diagram (\ref{kappaNvertex2}) 
since now the meson coupling to the nucleon can be either the $\pi$ or
the $K$, while in the case  of the p-wave, the requirement to include only
ordinary baryons in the intermediate baryon state does not allow the
$K$ to be coupled to the nucleon
\footnote{ Knowing the $\Theta^+$ existence, 
we could exchange this particle there. Yet, the intermediate 
state would be off-shell by about 170 MeV, plus the small KN width of the 
$\Theta^+$ makes the $KN\Theta^+$ coupling small, so this contribution can 
be safely neglected.}.

We need now the $m N \to m N$ amplitudes, which are easily obtained from 
the Lagrangian of eq. (\ref{lagrangian}) and give
\be \label{mN}
t_{mN\to mN} = - \frac{1}{4f^2} C_{ij}(q^0 + q^{0'})
\ee
where $q^0$, $q^{0'}$ are the initial, final meson energies and 
the $C_{ij}$ coefficients are given in table \ref{tablitacij}.
\begin{table} \caption{Flavor coefficients for meson-nucleon
scattering $C_{ij}$}\label{tablitacij}
\begin{ruledtabular}
\begin{tabular}{l|cccccccc}
$C_{ij}$  & $\pi^0 p$  &  $\pi^+n$  &  $\pi^-p$  &  $ \pi^0 n$ &
$K^0p$    & $K^+n$     & $K^+p$     & $ K^0n$ \\ \hline
$\pi^0 p$ & 0 & $\sqrt{2}$ & & & & & &        \\
$\pi^+n$  & $\sqrt{2}$ & 1 & & & & & &        \\
$\pi^-p$  & & & 1 & $-\sqrt{2}$ & & & &       \\
$ \pi^0n$ & & & $-\sqrt{2}$ & 0 & & & &       \\
$K^0p$    & & & & & $-1$ & $-1$ & &           \\
$K^+n$    & & & & & $-1$ & $-1$ & &           \\
$K^+p$    & & & & & & & $-2$ &                \\
$ K^0n$   & & & & & & & & $-2$                \\ \hline
\end{tabular}
\end{ruledtabular}
\end{table}

After performing the $q^0$ integration in the loop with three propagators 
with the explicit ($q^0+q^{0'}$) dependence of the vertex of eq. 
(\ref{mN}), but taking $t_{K \pi}$ with the arguments of the external 
$K\pi$ system, we obtain for the case of a $\pi$ coupling to 
the nucleon in (\ref{kappaNvertex3})
\be \label{mNsolved}
t_{mN}^{s'}= -2 t_{mm}^2 \frac{P^0}{4f^2} \int \frac{d^3q}{(2\pi)^3}
\frac{(\omega_\pi+\omega_K)}{\omega_\pi \omega_K} \left[ \frac{1}{P^{0 \ 
2}-(\omega_\pi + \omega_K)^2 } \right]^2 \ .
\ee
It is worth noting that this expression is symmetric in $\pi$ and 
$K$. Hence, the loops corresponding to having the K instead of the $\pi$ 
coupling to the
nucleon have the same expression up to some $SU(3)$ flavor factors. 
A straightforward calculation shows that for $I=0$ the coefficient 
is the same, but opposite in sign, whether the pion or the kaon couple to
the nucleon, 
implementing an exact cancellation of the two types of diagrams. 
It is also worth noting that in the case of I=1 there is no
cancellation but instead one finds a repulsive contribution,
obtained by changing the coefficient $2$ of eq. (\ref{mNsolved}) by 
$-\frac{2}{3}$.

Diagram (\ref{kappaNvertex3}), when the meson exchange is iterated between 
the other meson and the nucleon generates a subseries of the terms 
implicit in the Faddeev equations. 
For instance, the subseries of terms in 
the iterations of (\ref{kappaNvertex3})
with a pion generate the Faddeev series in the fixed center approximation,
accounting for the interaction of the pion with the $K N$ system (should 
it be bound by itself which is not the case) \cite{deloff,Kamalov:2000iy}.
Yet, this subseries is inoperative, given the cancellation of the $\pi$
and $K$ contributions.

Thinking along the same lines we are lead to the other subseries of the 
Faddeev equations in which the nucleon is the particle being
exchanged between the mesons:
\be \label{nucleonex}
\begin{picture}(200,50)(0,0)
\DashArrowLine(0,40)(200,40){7}
\DashArrowLine(0,0)(200,0){3}
\SetWidth{2}
\ArrowLine(0,20)(20,40) \ArrowLine(20,40)(60,0) 
\ArrowLine(60,0)(100,40) \ArrowLine(100,40)(140,0)
\ArrowLine(140,0)(180,40) \ArrowLine(180,40)(200,20)
\end{picture}
\ee
The basic vertex in this mechanism is 
\be \label{nucleonexvertex}
\ov{t}_{mB}^s = \ \ 
\begin{picture}(100,100)(-25,15)
\DashArrowLine(0,0)(30,40){3} \DashArrowLine(70,40)(100,0){3}
\DashArrowLine(30,70)(50,40){7} \DashArrowLine(50,40)(70,70){7}
\DashArrowArc(50,40)(20,-180,0){3}
\SetWidth{2}
\ArrowLine(0,40)(30,40) \ArrowLine(30,40)(50,40)
\ArrowLine(50,40)(70,40) \ArrowLine(70,40)(100,40) 
\end{picture}
\ee
where once again the upper meson can be a pion or a kaon.

Following the same techniques as before we obtain for this term's
contribution the result
\be \label{thirdvertex}
\ov{t}_{K N}^s+\ov{t}_{\pi N}^s = \frac{k^0 p^{0'}}{f^6} 
\left( p^{0'} \tilde{G}_{K\pi}(P^{0'}) - k^0 \tilde{G}_{\pi K}(P^0) 
\right)
\ee
with $P^0$ and $P^{0'}$ the energy of the kaon/nucleon and pion/nucleon 
pair respectively,
the loop function
\be \label{GpiK}
\tilde{G}_{\pi K}(P^0) = \int \frac{q^2 dq}{4\pi^2} \frac{1}{\omega_K}
\left( \frac{1}{P^0-\omega_K-E_N+i\epsilon} \right)^2
\left( \frac{m_N}{E_N} \right)^2
\ee
and $\tilde{G}_{K\pi}$ having the same expression permuting $\pi$ and $K$.
The contribution of
eq.(\ref{thirdvertex}) vanishes in the SU(3) limit of equal meson masses, 
but for unequal meson masses there is a net attractive contribution which 
has about the same strength as that of the four meson contact term of
eq. (\ref{diagtmatrix}).
Other interaction terms where the meson lines cross each other are 
possible, but either vanish like
\be
\begin{picture}(180,75)(0,0)
\SetWidth{2}
\ArrowLine(0,0)(45,0) \ArrowLine(45,0)(135,0) \ArrowLine(135,0)(180,0)
\SetWidth{0.5}
\DashArrowLine(6,75)(45,45){7} \DashArrowLine(45,45)(45,0){7} 
\DashLine(45,0)(135,45){7} \DashArrowLine(135,45)(171,75){7}  
\DashArrowLine(0,45)(45,45){3} \DashArrowLine(45,45)(86,23){3}
\DashArrowLine(94,17)(135,0){3} \DashArrowArc(90,21)(4,135,315){2}
\DashArrowLine(135,0)(135,45){3} \DashArrowLine(135,45)(180,45){3}  
\end{picture}
\ee
or are small since they involve baryons in the $t$-channel which are very
far off-shell such as
\be
\begin{picture}(180,75)(0,0)
\SetWidth{2}
\ArrowLine(0,0)(45,0) \ArrowLine(45,0)(135,0) \ArrowLine(135,0)(180,0)
\SetWidth{0.5}
\DashArrowLine(6,75)(45,45){7} \DashArrowLine(45,45)(45,0){7}
\DashLine(60,0)(135,45){7} \DashArrowLine(135,45)(171,75){7}
\DashArrowLine(0,45)(45,45){3} \DashArrowLine(45,45)(86,21){3}
\DashArrowLine(94,13)(120,0){3} \DashArrowArc(90,17)(3,135,315){2}
\DashArrowLine(135,0)(135,45){3} \DashArrowLine(135,45)(180,45){3}
\end{picture}
\ee
or involve one p-wave coupling inside a loop which makes it vanish for
large baryon mass, for example
\be
\begin{picture}(180,75)(0,0)
\SetWidth{2}
\ArrowLine(0,0)(45,0) \ArrowLine(45,0)(135,0) \ArrowLine(135,0)(180,0)
\SetWidth{0.5}
\DashArrowLine(6,75)(45,45){7} \DashArrowLine(45,45)(45,0){7}
\DashLine(135,0)(135,45){7} \DashArrowLine(135,45)(171,75){7}
\DashArrowLine(0,45)(45,45){3} \DashArrowLine(45,45)(90,0){3} 
\DashArrowLine(90,0)(135,45){3} \DashArrowLine(135,45)(180,45){3}
\end{picture}
\ \ .
\ee

\section{Bethe Salpeter iteration in the $(K \pi N)$ system}

Now we turn our attention to the formulation of the three body problem.
We have implemented the correlation between $\pi$ and $K$ through multiple
scattering, but we have not done so with the $K$ $N$ or $\pi$ $N$
interaction. In the case of the $K$ $N$ interaction this multiple
scattering barely changes the lowest order $t$ matrix $t_{mN\to mN}$ 
\cite{Oset:1997it}.
In the case of the $\pi$ $N$ system it generates attraction which is also
weak at the low energies considered here and only becomes sizeable around
$\sqrt{s}=1500 \ MeV$ where it leads, together with other coupled
channels, to the generation of the $N^*(1535)$ resonance
\cite{Kaiser:1995eg,Inoue:2001ip,InouepiN}.

The series of  $K \pi$  loop diagrams of (\ref{kappaprop}) is summed with 
the following equation;

\be
G^\kappa(s) = \frac{G_{mm}(s)}{1-t_{mm}(s) G_{mm}(s)} \ .
\ee
which yields a kappa propagator (that is, a propagator for a
correlated spin 1/2 pion-kaon state). 

At last, if the $\Theta$ was going to exist as a three-body bound
state, it should appear as a resonance of the $\kappa-N$ scattering matrix
which appears when summing the contribution of the diagrams of
(\ref{Thetaprop}), given by
\be \label{thetaLippman}
t_{\kappa N}(s)= \frac{t_{mB}(s)}{1-t_{mB}(s)G_{mB}(s)} 
\ee
where $t_{mB}$ sums the three non-vanishing contributions eqs.
(\ref{kappaNvertex1}, \ref{kappaNvertex2}, \ref{thirdvertex})
\be
t_{mB}=t_{mB}^s+t_{mB}^p+\ov{t}_{mB}^s \ .
\ee
The relevant loop function here, $G_{mB}$ appearing as the big loop in 
eq. (\ref{Thetaprop}), is made numerically more tractable by employing the
Lehmann representation for $G^\kappa$,
\be
G^\kappa (q^0,\vec{q}) = \frac{-1}{\pi} \int_{m_\pi+m_K}^\infty d\omega
\frac{2\omega {\rm Im}G^\kappa(\omega^2-\ar \vec{q}\ar^2)}{q_0^2-\omega^2}
\ .
\ee
(although we have checked our codes also by direct computation).
After factorizing the vertices with the on-shell
prescription, we obtain
\ba 
G_{mB}(s) =
\frac{-1}{2\pi^3}\int_0^\Lambda q^2dq \frac{M_N}{E_N(q)}
\int_{\sqrt{(m_\pi+m_K)^2+q^2}}^\infty d\omega
\frac{{\rm Im} G^\kappa(\omega^2-q^2)}{\sqrt{s}-\omega-E_N(q)}
\\ \nonumber\\ \nonumber
{\rm with}\ \ \   \sqrt{s}\ \in \ (m_N+m_K,m_N+m_K+m_\pi)
\ .
\ea
The algebraic formulation of the Bethe Salpeter eq. (\ref{thetaLippman})
is possible because we have factorized the $(k^0+k'^0)$, $(p^0+p^{0'})$
dependence of eq. (\ref{diagtmatrix}, \ref{supertmb2}, \ref{thirdvertex}) 
with its on-shell value given by the
external variables.
We have
performed the loop integrals with the full off shell part and found
that the on shell approximation induces errors of less than 20 per cent,
hence, it is accurate enough for the exploratory purpose of the present
work.

There is a technical detail worth mentioning. We have assumed in the
calculations that the incoming and outgoing particles have zero momentum.
This is certainly an approximation, but it simplifies the calculations
since in the diagram (\ref{kappaNvertex2}) one has two identical pions
propagating and in (\ref{nucleonexvertex}) one has two identical nucleon
propagators and we evaluate these Feynman diagrams by partial derivation
of a loop function with only one pion
or one nucleon propagator respectively.  This causes no problem if one
investigates the amplitudes at 30 $MeV$ below threshold but the
approximation induces an infrared divergence at threshold. We are not
interested in this region but in any case we cure the divergence by
assuming an average momentum of the particles in the three body wave
function. We take $100 \ MeV/c$ for this momentum and we should change
$\omega_\pi(\vec{q})$ by $\omega_\pi(\vec{q}-\vec{p})$ which close 
to threshold can be approximated by $\omega_\pi(q) + \frac{p^2}{2m_\pi}$ .
Similarly, for the diagram (\ref{nucleonexvertex}), the nucleon energy is
changed to $E_N(q)+\frac{p^2}{2m_N}$. This cures the infrared divergence
at threshold and has negligible influence away from it.

\section{Numerical Results and Conclusion.} \label{results}
\begin{figure}[h] 
\psfig{figure=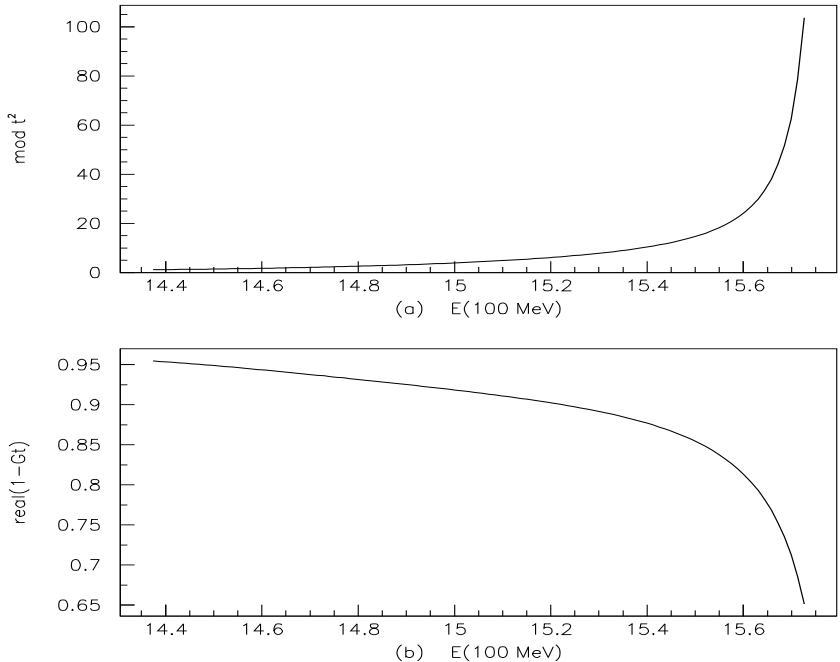,width=5in,height=4in}
\caption{\label{amplitude} Our final result: $K\pi N$
scattering matrix (modulus and denominator of eq. (\ref{thetaLippman})).
Energy units are $100 \ MeV$.}
\end{figure}

\begin{figure}[h] 
\psfig{figure=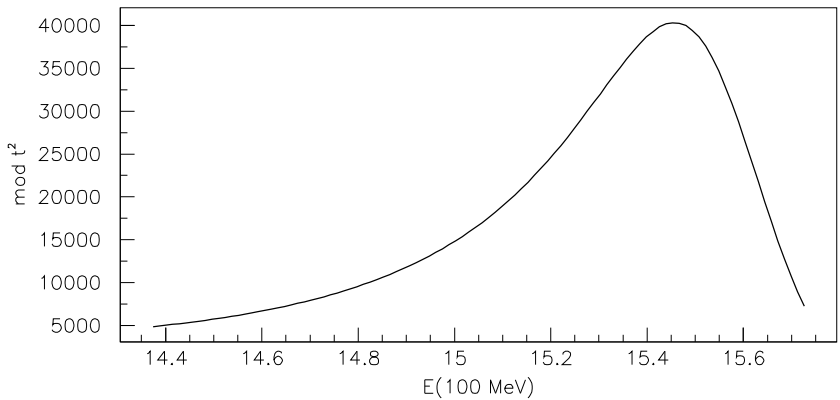,width=5in,height=5in}
\caption{\label{trickedamplitude1} We find a resonance
with a reasonable width for a potential larger by a factor 6 (see text).
Units are $100 \ MeV$.}
\end{figure}

\begin{figure}[h] 
\psfig{figure=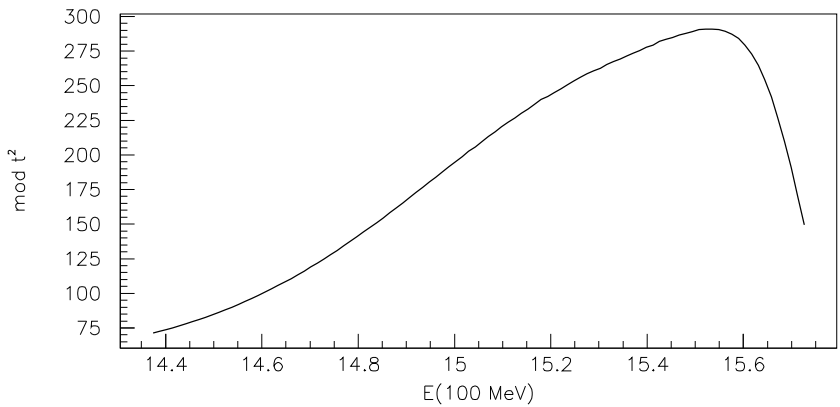,width=5in,height=5in} 
\caption{\label{trickedamplitude2} We also find a
resonance, this time too broad, by increasing the cutoff in the $\kappa
N$ loop to 4 $GeV$ and about doubling the potential. Units are $100 \
MeV$.}
\end{figure}

We examine now the $t_{\kappa N}$ amplitude of eq. (\ref{thetaLippman})
as a function of $\sqrt{s}$ of the three external particles (for 
simplicity we split the small binding energy between the pion and kaon in 
proportion to their masses). In figure
\ref{amplitude} (a) we show $\ar t\ar^2$ against $\sqrt{s}$.
We see that the function is monotonously increasing as a function
of $\sqrt{s}$, but there is no trace of a pole or resonance. In order to
see how far we are from a pole, we show in fig. \ref{amplitude} (b)
the real part of the denominator of eq. (\ref{thetaLippman}),
$1-t_{mB}G_{mB}$. We see that in the region from $\sqrt{s}=1540 \ MeV$
 to $1570 \ MeV$ this value is bigger than 0.6, while it should be around
zero to have a resonance. Typical values of $t_{mB}$ and $G_{mB}$ are
$G_{mB}\simeq -0.05 \ (100 \ MeV)^3$, $t_{mB}\simeq -(2-3) \ (100 \
MeV)^{-3}$ for a cutoff $\Lambda=1 \ GeV$. From these results we can
conclude that
\begin{enumerate}
\item With the dynamics which we are considering we find no bound state
around $\sqrt{s}=1540 \ MeV$.
\item The fact that $t_{mB}G_{mB}$ is far away from unity indicates that
we are far away from having a pole of the $\kappa N$ scattering matrix.
\end{enumerate}
In order to quantify this second statement we proceed as follows.
We increase artificially the potential $t_{mB}$ by adding to it a quantity
which leads to a pole around $\sqrt{s}=1540 \ MeV$. This is reached by
adding $-16 \ (100 \ MeV)^{-3}$ to the already existing potential, which
means we add an attractive potential around five or six times bigger than
the existing one. If we do that we obtain the results for $\ar t \ar ^2$
shown in figure \ref{trickedamplitude1}.
There is indeed a resonance around $\sqrt{s}=1540 \ MeV$ with a width of
around $\Gamma=40 \ MeV$, which is of the order of magnitude of the
experimental one. Refinements of the theory considering that in the
generation of the resonace the external $\kappa$ would be itself part of a
loop, would lead according to our estimates to a smaller width, but for 
the order of magnitude the approximations performed are fair.
This exercise gives a quantitative idea of how far one is from having a
pole. We do not envisage at this stage a possible source of such 
a large attraction within our theoretical treatment.
There is another exercise which we want to present here. We have
regularized the $\pi K N$ loop function with a cutoff in the three
momentum of $1 \ GeV$. This is the natural scale for the problems we are
dealing with. Yet, we could try to see how much $\Lambda$ has to be
increased to find a pole. The exercise conducted is the following:
we have taken $\Lambda=4 \ GeV$ and see how much more potential we have to
add to get the resonance around $1540 \ MeV$. This is done by adding a
potential with a strength of $-2.5 \ (100 \ MeV)^{-3}$, which amounts to
about doubling the calculated one. The result for $\ar t \ar^2$ 
can be seen in fig. \ref{trickedamplitude2}.
What we see is that the width becomes much larger than before. 
This trend continues in the same manner and we can reduce the amount
of extra potential as the cutoff $\Lambda$ increases (although the
dependence of $G$ on the cutoff is by then logarithmic). The width also
increases unrealistically for these larger values of $\Lambda$. Hence 
this does not seem to be the adequate path to follow in future searches.

As a positive output there are hopes, given by the trend of the results in
fig. \ref{amplitude} that a resonance could develop at higher energies
above threshold. This would be a task worth following that however would
require to modify technically our approach which has relied on a below
threshold situation avoiding the singularities of open physical
channels above threshold.

Another point is that we have only partially solved the Faddeev equations,
including therein  a three body potential  with the basic units
repeated in the Faddeev sequence of diagrams. A more standard three body 
Faddeev approach would also be one of the tasks worth undertaking. The
steps walked here and the dynamics used could be directly  input to 
the full set of Faddeev equations.

In summary, we think our calculation is sufficiently accurate to claim
that the nature of the $\Theta^+$ as  a bound $\pi K N$ system is very
unlikely, but this should be checked by other independent calculations
and different technical approaches given the importance of this resonance.
At last, it would also be interesting to continue with the
present study extrapolating the approach above $K \pi N$ threshold to
explore the possibility of a resonance at not too high energies but beyond
the scope of the present work.

\acknowledgments
One of us, E. O., acknowledges the hospitality of the Universidad 
Complutense de Madrid where most of the work was done and useful 
discussions with T. Nakano, J. Nieves and J. A. Oller.   
V. M.   acknowledges CSIC financial support within the program of summer 
stays in the IFIC at Valencia.
This work is also supported in part  by DGICYT projects BFM2000-1326,
FPA 2000-0956, BFM 2002-01003 (Spain)
and the EU network EURIDICE contract
HPRN-CT-2002-00311.



\end{document}